\documentclass[a4paper,twocolumn,11pt]{quantumarticle}
\pdfoutput=1
\usepackage{amsmath}
\usepackage{graphicx}
\usepackage{dcolumn}
\usepackage[numbers,sort&compress]{natbib}
\usepackage{bm}
\begin{document}

\title{Direct detection of quantum non-Gaussian light from a dispersively coupled single atom }

\author{Jitendra K. Verma}
\thanks{jitendrakumar.verma@upol.cz}
\author{ Luk\' a\v s Lachman}
\thanks{lachman@optics.upol.cz}
\author{Radim Filip}%
\thanks{filip@optics.upol.cz}
\affiliation{%
 Department of Optics, Faculty of Science, Palack\' y University, 17. listopadu 1192/12,  771~46 Olomouc, Czech Republic}

\begin{abstract}
 Many applications in quantum communication, sensing and computation need provably
quantum non-Gaussian light. Recently such light, witnessed by a negative Wigner function, has
been estimated using homodyne tomography from a single atom dispersively coupled to a high-finesse cavity \cite{Hacker2019}. This opens an investigation of quantum non-Gaussian light for many experiments with atoms and solid-state emitters.
However, at their early stage, an atom or emitter in a cavity system with different channels to the environment and additional noise are insufficient to produce negative Wigner functions. Moreover, homodyne detection is frequently challenging for such experiments. We analyse these issues and prove that such cavities can be used to emit quantum non-Gaussian light employing single-photon detection in the Hanbury Brown and Twiss configuration and quantum non-Gaussianity criteria suitable for this measurement. We investigate in detail cases of considerable cavity leakage
when the negativity of the Wigner function disappears completely. Advantageously, quantum non-Gaussian light can be still conclusively proven for a large set of the cavity parameters at the cost of overall measurement time, even if noise is present. 
\end{abstract}

\maketitle

\section{Introduction}
Discrete electronic level structures in the individual atoms and solid-state systems embedded in a cavity are excellent candidates for future deterministic and scalable light sources for quantum technology \cite{OextquotesingleBrien2009, Aharonovich2016}. Such emitted light is frequently nonclassical, however, quantum communication \cite{Wang2021}, quantum sensing \cite{Wolf2019}, quantum simulations \cite{AspuruGuzik2012} and quantum computing \cite{Duan2004} require light beyond the Gaussian squeezed light produced in nonlinear optics. The literature primary focuses on high quality and scalable generation of single-photon states \cite{Lodahl2015,Arcari2014,Somaschi2016,Morin2019,Zapletal2017,Huh2015}. 


However, current quantum technology already requires a broader class of sources than ones repeatedly producing single-photon states \cite{Lodahl2004,Peter2005,Chu2016,Ding2016} or photon-pairs \cite{Prilmueller2018,Gines2021} on demand. Encoding information through the large dimensional space of continuous variables needs some form of coherent-state superposition instead \cite{Fluehmann2019}. At optical frequencies, only conditional generation of such states has been available so far \cite{Wakui2007}. However, that method is not scalable, and it seriously limits optical continuous variable experiments. Atomic and solid-state cavity experiments have the potential to be deterministic and scalable sources of such quantum non-Gaussian states \cite{Reiserer2015,Thomas2021}. Therefore, the generation of a coherent-state superposition from a single atom dispersively coupled to an optical cavity was a pioneering step \cite{Hacker2019}.  This superposition exhibited highly nonclassical negative values of the Wigner function, indirectly estimated from homodyne tomography \cite{Lvovsky2009}.  This opens new territory for atomic and solid-state experiments that could finally cover demands for hybrid quantum optical technologies and their applications \cite{Andersen2015}. However, initially, the cavity development and overall loss are still limiting factors for many experiments to observe such negative Wigner functions \cite{Huber2017}. This then limits the transition from routinely detected nonclassical features of such light to more challenging quantum non-Gaussian aspects. While Wigner function negativity provides a conclusive proof that light is beyond any mixture of Gaussian states it is unnecessarily high threshold \cite{Hudson1974}. In addition, homodyne tomography is too demanding for many experiments due to the requirement of mode matching with the local oscillator. 

In this paper, we propose direct detection of quantum non-Gaussian light from a single atom in the cavity, which employs single-photon avalanche photo-diodes (SPADs) in Hanbury Brown and Twiss configuration instead of homodyne tomography. Importantly, this method allows us to detect quantum non-Gaussian light from a single atom in the leaky cavity limit when the negative Wigner function is not observable. This enables investigation of quantum non-Gaussian features produced by discrete levels in a broad class of current atomic and solid-state experiments. Such experiments will further open a general theoretical and experimental investigation of such light sources and stimulate further steps in this part of photonics and quantum technology. 

\begin{figure*}[ht!]
\centerline {\includegraphics[width=0.9\linewidth]{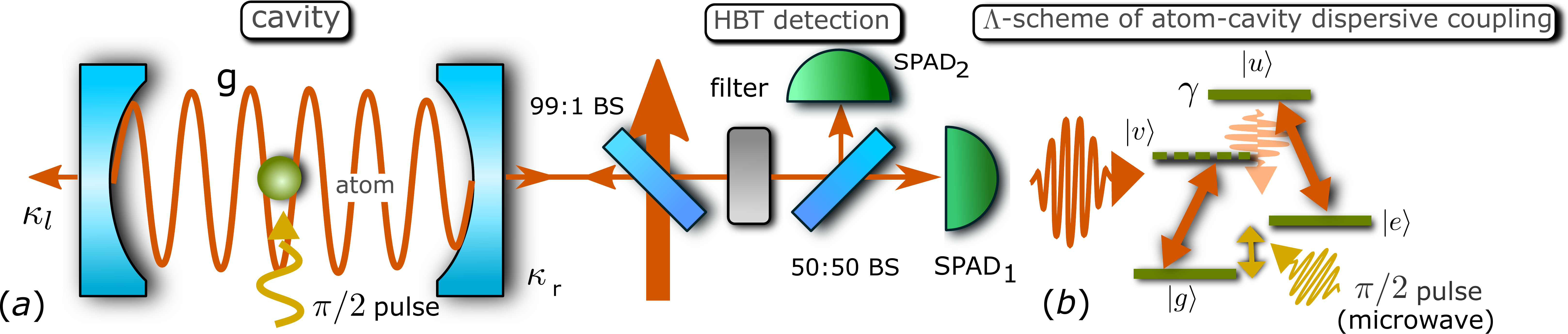}}
\caption{Quantum non-Gaussian light from a single atom in a leaky cavity. ({\emph a}) A scheme for preparation and detection of quantum non-Gaussian light from a cavity. A pulsed laser beam (orange) goes through a BS with very low reflectivity and a small portion of the beam is reflected towards a cavity where it interacts dispersively with an atom (see ({\emph b})). The mirrors of the cavity exhibit losses $\kappa_l$ and $\kappa_r$. The state of the atom is simultaneously manipulated by a microwave field during the interaction. The light leaving the cavity is measured in the Hanbury Brown and Twiss detection set-up. It comprises a beam-splitter (BS), which divides the impinging state of light between two single-photon avalanche photo-diodes SPAD$_1$ and SPAD$_2$. ({\emph b}) A scheme of the employed energy transitions forming the $\Lambda$-system. The laser pulse (orange wave) entering the cavity drives the transition between states $|e\rangle$ and $|u\rangle$ resonantly and $|g\rangle$ and $|u\rangle$  off-resonantly through the virtual level $|v\rangle$. The state $|u\rangle$ can decay spontaneously (light orange wave) with the rate $\gamma$. The initial atomic state $(|g\rangle+|e\rangle)/\sqrt{2}$ causes entanglement between the light emerging from the cavity and the atom. Simultaneous interaction of the atom with a $\pi/2$ microwave pulse (yellow wave) addressing the transition between $|g\rangle$ and $|e\rangle$ yields the state (\ref{catDM}) that can exhibit heralded quantum non-Gaussian light by measuring the atomic state.}
\label{fig:fig1}
\end{figure*}

\section{Quantum non-Gaussianity}
Quantum non-Gaussianity refers to overcoming any limitation given by the mixtures of states $|\psi_G\rangle = S(\xi)D(\alpha)|0\rangle$, where the squeezing operator $S(\xi)=\exp\left[\xi \left(a^{\dagger}\right)^2-\xi^* a^2\right]$ and the displacement operator $D(\alpha)=\exp(\alpha a^{\dagger}-\alpha^* a)$ acts on the vacuum. Therefore, the density matrix $\rho$ exhibits quantum non-Gaussianity formally when
\begin{equation}
    \rho \neq \int \mathrm{d}^2 \xi \mathrm{d}^2\alpha P(\xi,\alpha) S(\xi)D(\alpha) \vert 0 \rangle \langle 0 \vert D^{\dagger}(\alpha) S^{\dagger}(\xi),
    \label{DefQNG}
\end{equation}
where $P(\xi,\alpha)$ stands for a probability density function of the parameters $\xi$ and $\alpha$. 
 Quantum non-Gaussianity can be recognized under realistic conditions of current experiments employing the Hanbury Brown and Twiss detection scheme \cite{Brown1954}, which is depicted in Fig.~\ref{fig:fig1} (\emph{a}). It consists of a beam-splitter (BS) and two single-photon avalanche photo-diodes SPAD$_1$ and SPAD$_2$. A criterion \cite{Filip2011, Lachman2013} for the quantum non-Gaussianity compares their response with a threshold defined in terms of success probability $P_s$ of a click on SPAD$_1$ and error probability $P_e$ of clicks on both detectors SPAD$_1$ and SPAD$_2$. The quantum non-Gaussianity manifests itself when a state exhibits the probabilities $P_s$ and $P_e$ that surpass a threshold given by the parametric expression
\begin{equation}
\begin{aligned}
 P_s(t)&=1-4e^{-\frac{1-t^2}{2 t(1+3t)}}\sqrt{\frac{t}{3t^2+10 t+3}}\\
P_e(t)&=1-8e^{-\frac{1-t^2}{2 t(1+3t)}}\sqrt{\frac{t}{3t^2+10 t+3}}\\
&+e^{-\frac{(t-1)(t+3)}{2 t(1+3t)}}\frac{\sqrt{t}}{1+t},
\end{aligned}
\label{PsPeTh}
 \end{equation}
where $t\in\left(0,1\right]$ corresponds to a parameter identifying the tuple $(P_s,P_e)$ giving the threshold values. Surpassing the threshold (\ref{PsPeTh}) means a measured state achieves success probability $P_s$ greater for a given error probability $P_e$ than any mixture of Gaussian states allows.
Alternatively, modern experimental platforms exploit homodyne tomography to perform the detection. It relies on the difficult matching of a local oscillator to the output mode of a cavity with an atom \cite{Hacker2019} and can be used to estimate the density matrix in the Fock state basis or the Wigner function in phase space.
 With the knowledge of the density matrix, the quantum non-Gaussianity manifests itself already in two probabilities $p_1=\langle 1 \vert \rho \vert 1 \rangle$ and $p_{2+}=1-\langle 0 \vert \rho \vert 0 \rangle-\langle 1 \vert \rho \vert 1 \rangle$  by exceeding the threshold
\begin{equation}
    p_1=2e^{({\frac{t-1}{2t}})}\frac{1-t}{\sqrt{t}(1+t)^2}; p_{2+}=1-2e^{({\frac{t-1}{2t}})}\frac{1+t^2}{\sqrt{t}(1+t)^2},
    \label{PsPeThPNRD}
\end{equation}
where $t\in \left(0,1\right]$ parmetrises the threshold again. This criterion is also suitable for partial photon-resolving detectors faithfully distinguishing zero, single and more photons.

Using homodyne tomography, quantum non-Gaussianity is certified by the negativity of the Wigner function \cite{Lvovsky2009}. However, the negativity always vanishes for $3$ dB of loss, which is too challenging for many experimental platforms in optics \cite{Aharonovich2016}. For large losses, certification of the quantum non-Gaussianity employing the Wigner function becomes less demanding on the losses if a constraint on the mean number of photons is involved in the criterion \cite{Palma2014}. However, this requires either quantum tomography using homodyne detection, challenging for many atomic and solid-state experiments or, ideally, direct detection resolving all the photon numbers to determine the mean number of photons without systematic errors. on the contrary, criterion (\ref{PsPeTh}) uses only the most common method of photon auto-correlation measurement and criterion  (\ref{PsPeThPNRD}) employs only detectors distinguishing no photons, single photon and more than one photon. Compared to \cite{Palma2014}, this largely relaxes the conditions for conclusive and faithful photon number resolution. Simultaneously, they are more demanding on the quality of prepared states than the nonclassicality criterion \cite{Grangier1986,Lachman2013}, rejecting all classical waves. In spite of that, it was demonstrated both theoretically \cite{Lachman2016} and experimentally that criteria (\ref{PsPeTh}) and (\ref{PsPeThPNRD}) are feasible for solid-state sources \cite{Straka2014} and atomic sources \cite{Higginbottom2016} when the negativity of the Wigner function disappears completely due to optical losses. The quantum non-Gaussianity of such states was recognized as a useful indicator for safety in quantum communication \cite{Lasota2017}. However, no test has been done yet for the atomic or solid-state emitters in the cavity.



\section{Single atom dispersively coupled to a cavity} 
A prospective platform that enables diverse generation and detection of the quantum non-Gaussian light from a single atom coupled dispersively to a cavity is depicted in Fig.~\ref{fig:fig1} (\emph{a}). The BS with very small reflectivity splits a weak coherent state from the pumping beam and the coherent state enters the cavity mediating the interaction between light in the cavity mode and an atom.  Fig.~\ref{fig:fig1} (\emph{b}) presents relevant atomic energy levels forming the $\Lambda$-scheme. Whereas the transition between states $|e\rangle$ and $|u\rangle$ is coupled to the cavity mode, the transition between states $|g\rangle$ and $|u\rangle$ is strongly detuned from the cavity mode. A cavity operating in the strong coupling regime \cite{Reiserer2015} induces dispersive coupling between the atom and light inside the cavity, which allows the atom to control the phase of the coherent state leaving the cavity \cite{Duan2004}. If the atom is in the state $|g\rangle$, the cavity acts as a mirror due to off-resonant coupling with cavity mode and, in this case, the coherent state is reflected with a $\pi$ phase shift. When the atom occupies the state $|e\rangle$, the strong coupling with the cavity mode induces significant detuning of the cavity dressed states from the frequency of the incoming coherent state. Thus, the reflected state remains a coherent state, which experiences no phase shift \cite{Wang2005,Goto2005}. Preparing the atom initially in the superposition $(|g\rangle+|e\rangle)/\sqrt{2}$ gives rise to entanglement between light leaving the cavity and the atomic state \cite{Hacker2019}. Employing simultaneously a microwave pulse that causes $\pi/2$ rotation between the states $|g\rangle$ and $|e\rangle$ allows us to generate the state
\begin{equation}
\begin{aligned}
    |\psi \rangle &=\frac{1}{\sqrt{N(\alpha)}}\left[( \vert \alpha \rangle-\vert -\alpha \rangle)|e\rangle \right.\\
    &\left. +( \vert \alpha \rangle+\vert -\alpha \rangle)|g\rangle \right],
\end{aligned}
\end{equation}
where $\alpha$ is the amplitude of the coherent state entering the cavity and $N(\alpha)$ is such that $\langle \psi |\psi \rangle=1$ \cite{Daiss2019}. Thus, projection to the atomic state $|e\rangle$ heralds the state of light
\begin{equation}
\vert \psi_{-} \rangle = \frac{1}{\sqrt{2-2 \exp(-\vert \alpha \vert^2)}}( \vert \alpha \rangle-\vert -\alpha \rangle),
\label{catSt}
\end{equation}
where the superposition of coherent states with different phase has the capacity to exhibit quantum non-Gaussianity due to negative Wigner function, in principle, for any $\alpha$.

\begin{figure*}[ht!]
\centering
\includegraphics[width=0.95\linewidth]{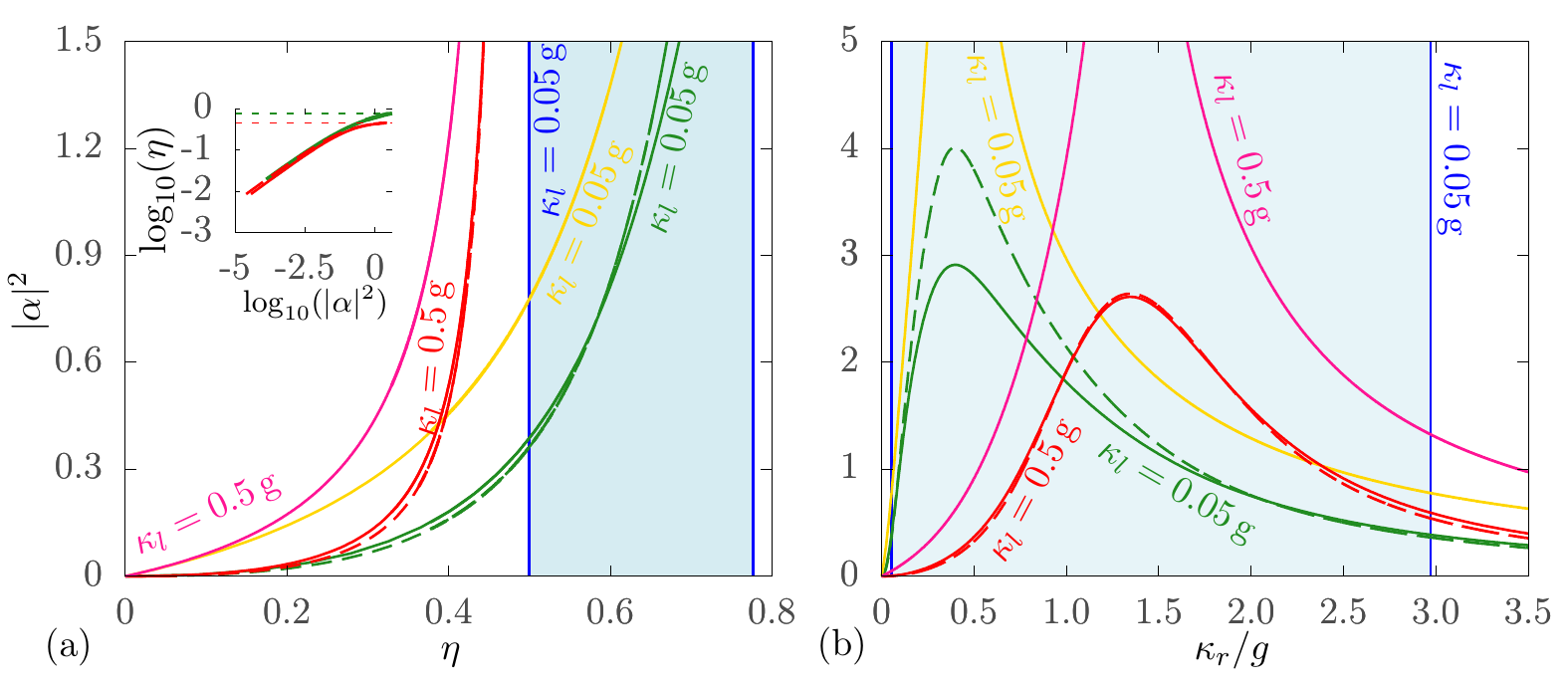}
    \caption{Quantum non-Gaussianity of light emitted by a single atom in a cavity in terms of the parameters of the dispersive interaction. The quantum non-Gaussianity manifests itself up to some threshold intensity $|\alpha|^2$ of the incident coherent light that is depicted against the cumulative parameter $\eta$ (\emph{a}) and cavity output decay rate $\kappa_r$ normalized to atom-cavity coupling rate $g$ (\emph{b}). In both figures, the solid and dashed lines represent the thresholds $|\alpha|^2$ varying with $\kappa_r$ for fixed spontaneous emission rate $\gamma=0.32 g$ and cavity damping rate either $\kappa_l=0.05 g$ or $\kappa_l=0.5 g$. The colors differentiate values of $\kappa_l$. Whereas the red and green solid lines present the threshold imposed by criterion (\ref{PsPeTh}), employing responses from SPADs, the same lines in dashed show the threshold using criterion (\ref{PsPeThPNRD}) derived for homodyne detection. The red and green dashed vertical lines in (\emph{a}) represent the physical boundary for the parameter $\eta$ when $\kappa_l=0.5 g$ and $\kappa_l=0.05 g$ respectively. In both figures, the thresholds for the cavity with damping rate $\kappa_l=0.05 g$ (green lines) are compared with the boundary for the negativity in the Wigner function (the light blue region). The cavity with damping rate $\kappa_l=0.5 g$ does not enable observation of the negativity. A condition imposed by the criterion of nonclassicality $P_s^2-P_e>0$ employing SPADs \cite{Lachman2013} is given by the yellow and pink solid lines. The inset plot in (\emph{a}) zooms the region of small $|\alpha|^2$ and $\eta$ in logarithmic scale. Briefly, (\emph{a}) demonstrates that the quantum non-Gaussianity can be detected for much smaller $\eta$ than the negativity of the Wigner function if the pumping power is low. In contrast, (b) shows that varying $\kappa_r$ strongly affects the maximal intensity $|\alpha|^2$ allowing the quantum non-Gaussianity to be exposed.}
\label{fig:fig2}
\end{figure*}

A realistic platform suffers from optical leakage from the cavity and dissipating emission from the atom. Simultaneously, the temporal shape of the reflected optical pulse gets distorted from the entering pulse when the amplitude $\alpha$ becomes large \cite{Wang2005}. Both these imperfections limit recognition of the quantum non-Gaussianity not only for atoms but also for solid-state systems. Further, we consider realistic states induced by the incident optical pulse with small amplitude $\alpha$. Then, the pulse distortion can be neglected and the quantum non-Gaussianity gets lost mainly due to  the leakage from the cavity or dissipating emission from the atom. Let $g$ denote the atom-cavity coupling rate between the transition $|e\rangle$-$|u\rangle$ and the cavity mode. The excitation to the atomic state $\vert u \rangle$ can cause a photon to be spontaneously emitted with rate $\gamma$. Also, the mirrors effects leakage of the light from the cavity towards the detector and towards the environment, which happens with rates $\kappa_r$ and $\kappa_l$, respectively. Considering these processes in the dynamics of the light coupled to the atom allows us to establish the approximate density matrix of the heralded state emerging from the cavity as \cite{Wang2005}
\begin{equation}
\begin{aligned}
    &\rho_{-}=\frac{1}{2(1-e^{-2\eta \vert \alpha \vert^2})}\left[\vert \alpha_e \rangle \langle \alpha_e \vert+\vert \alpha_g \rangle \langle \alpha_g \vert \right.\\
    &\ \left. -e^{-2(1-\eta)\eta \vert \alpha \vert^2}\left(\vert \alpha_e \rangle \langle \alpha_g \vert+\vert \alpha_g \rangle \langle \alpha_e \vert\right)\right],
\end{aligned}
\label{catDM}
\end{equation}
where the amplitudes of the coherent states $\vert \alpha_e \rangle$ and $\vert \alpha_g \rangle$ obey
\begin{equation}
    \alpha_g=\frac{\kappa_l-\kappa_r}{\kappa_l+\kappa_r} \alpha, \  \alpha_e=\frac{g^2+(\kappa_l-\kappa_r)\gamma}{g^2+(\kappa_l+\kappa_r)\gamma}\alpha.
    \label{alfGalfE}
\end{equation}
The cumulative parameter $\eta$ in (\ref{catDM}) reads
\begin{equation}
    \eta=\frac{\kappa_r}{\kappa_l+\kappa_r}\frac{g^2}{g^2+\gamma (\kappa_l+\kappa_r)}
    \label{eta}
\end{equation}
and it determines the purity of the state (\ref{catDM}). The strong-coupling regime \cite{Reiserer2015} signifies $1-\eta \ll 1$, for which $\rho_-$ approaches (\ref{catSt}) asymptotically.
 According to (\ref{eta}), the cavity has inherent imperfections quantified by $\kappa_l$ and $\gamma$, which reduce $\eta$ beyond the strong coupling regime. The parameter $\kappa_r$ has non-trivial impacts on $\eta$. Very small $\kappa_r$ causes the emission of the light towards the SPADs to be weak since the light is kept inside the cavity. On the other hand, high $\kappa_r$ increases the intensity of the light impinging on the SPADs but reduces the parameter $\eta$. With fixed $\gamma>0$ and $\kappa_l>0$, the maximal $\eta$ occurs for \cite{Goto2019}
 \begin{equation}
    \kappa_{r}=\sqrt{\frac{\kappa_l}{\gamma}}\sqrt{g^2+\gamma \kappa_l},
    \label{kappar}
\end{equation}
which expresses how the cavity can be engineered to maximize purity of the state $\rho_-$.


The state (\ref{catSt}) exhibits a Wigner function with fringes of negative regions, which proves the quantum non-Gaussianity.  However, the negativity is sensitive to optical leakage from the cavity and dissipative emission from the atom. Thus, the state $\rho_-$ manifests negative Wigner function only for $\eta>1/2$ (see Appendix for details concerning derivation). Considering $\kappa_r$ in (\ref{kappar}), the other parameters allow us to detect the negative Wigner function only when
 \begin{equation}
     \kappa_{l} \gamma< \frac{g^2}{8}.
     \label{negWig}
 \end{equation}
This shows that the product $\kappa_{l} \gamma$ is a key parameter for the exhibition of the negativity in the Wigner function. It implies that the negativity requires sufficient suppression of either $\gamma$ or $\kappa_l$ (ideally both) with respect to small $g$ and $\kappa_r$ approaching the optimal one given by (\ref{kappar}). 
 
 

The threshold (\ref{PsPeTh}) employing a response from SPADs can be surpassed even when condition (\ref{negWig}), which is necessary for the negativity of the Wigner function, fails. We evaluated the manifestation of the quantum non-Gaussianity by using the criterion that the threshold (\ref{PsPeTh}) yields. It imposes a condition on the parameter $\eta$ that varies with the intensity $\vert \alpha \vert^2$ of the incident coherent state. Fig.~\ref{fig:fig2} presents the results of this numerical analysis for two different states (\ref{catDM}). One of the states is given by $\gamma=0.32 g$ and $\kappa_l=0.05 g$, which represent the parameters of the performed experiment in \cite{Hacker2019}. These parameters fulfill condition (\ref{negWig}) allowing us to observe the negativity of the Wigner function for a proper choice of $\kappa_r$. The other state emerges from the cavity with $\kappa_l=0.5 g$ and $\gamma=0.32 g$. Due to (\ref{negWig}), its Wigner function is positive for any $\kappa_r$ and $|\alpha|^2>0$. Fig.~\ref{fig:fig2} (a) presents a condition the quantum non-Gaussianity imposes on the cumulative parameter $\eta$ with respect to the intensity $|\alpha|^2$ for the fixed parameters $\kappa_l$ and $\gamma$. In both considered states, the criterion (\ref{PsPeTh}) reveals the quantum non-Gaussianity even for situations when the Wigner function is completely positive as the figure demonstrates. Although the parameter $\eta$ is convenient for identifying the negativity of the Wigner function, its quantification of the cavity for the quantum non-Gaussianity is not so intelligible. Thus, the same results are presented in Fig.~\ref{fig:fig2} (b) in terms of $\kappa_r$ (normalized to fixed $g$) instead of $\eta$. It clearly demonstrates $\kappa_r$ reaching the value in (\ref{kappar}) determines the maximal $\vert \alpha \vert^2$ that allows the state (\ref{catDM}) to fulfill the criterion. The size of $\vert \alpha \vert^2$ becomes crucial in an analysis of the experimental error bars as described further. Also, both Figs.~\ref{fig:fig2} (a) and \ref{fig:fig2} (b) present the difference in the quantum non-Gaussianity manifestation employing the detection scheme with SPADs and homodyne tomography according to the criteria (\ref{PsPeTh}) and (\ref{PsPeThPNRD}) respectively. Both requirements become almost identical for states with $|\alpha|^2 \ll 1$. Beyond that regime, the conditions diversify themselves substantially only for the cavity obeying (\ref{negWig}), in which case the criterion (\ref{PsPeTh}) employing SPADs tolerates higher $|\alpha|^2$. Finally, these two figures compare the conditions for the quantum non-Gaussianity with the nonclassical condition that is less demanding on $\eta$ for a given $|\alpha|^2$ than quantum non-Gaussianity.

\begin{figure}[t]
\centering
\includegraphics[width=1.0\linewidth]{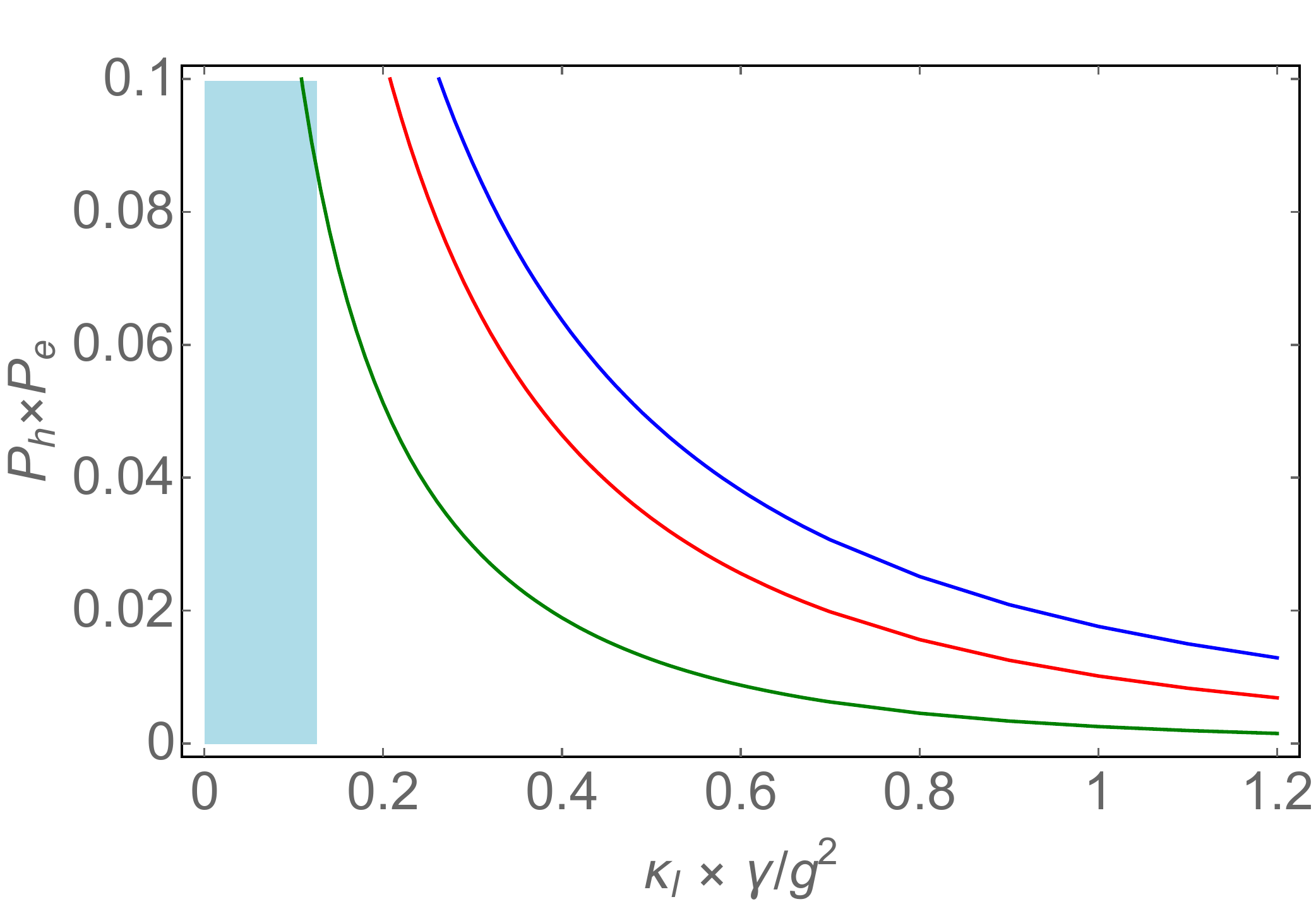}
\caption{The maximal probability $P_h \times P_e$ that allows recognition of the quantum non-Gaussianity against the product of parameters $\kappa_l\times \gamma/g^2$ and $k_r$ achieving the value $\kappa_{r,0}$ from (\ref{kappar})  (blue), $75\%$ of $\kappa_{r,0}$ (red) and $50 \%$ of $\kappa_{r,0}$ (green). The vertical axis quantifies the probability $P_h \times P_e$ exhibited by the states where the intensity $|\alpha|^2$ achieves its maximal value that allows certification of the quantum non-Gaussianity. Further growth of  $P_h\times P_e$ effected by increasing the intensity $|\alpha |^2$ of the pumping beam causes the failure in the recognition. The light blue stripe depicts cases with $\kappa_l \gamma/g^2<1/8$, which enables detection of the negativity in the Wigner function when $\kappa_r$ obeys (\ref{kappar}).}
\label{fig:fig4}
\end{figure}

Criterion (\ref{PsPeTh}) imposes no fundamental limit on the parameter $\eta$, i.e. any $\eta>0$ allows us to observe the quantum non-Gaussianity if $|\alpha |^2$ is sufficiently small. However, the intensity $\vert \alpha \vert^2$ influences the experimental time in recognition of the quantum non-Gaussianity since uncertainty in experimental estimation of the probabilities $P_s$ and $P_e$ depends on a number of click events, which grows with $|\alpha|^2$. Simultaneously, $|\alpha|^2$ affects the probability $P_h$ of successfully heralding the state $\rho_-$, which is given by projection on the atomic state $|e\rangle$, according to $P_h=1-e^{-2\eta |\alpha|^2}$. This suggests the appropriate intensity $|\alpha|^2$ is given by the trade-off between the overall measurement time and requirement imposed by the suppression of the threshold (\ref{PsPeTh}). Fig.~\ref{fig:fig4} presents the rate $P_h \times P_e$ for states achieved from the maximal intensity $|\alpha |^2$ of the pumping beam that the quantum non-Gaussianity tolerates. Note, the rate is always lower in real detection of the quantum non-Gaussianity since it has to be arranged by $|\alpha|^2$ such that the measured probabilities $P_s$ and $P_e$ surpass the threshold even within their error bars.

\section{Realistic cavity with noise contributions}

Realistic systems can be further affected by small noise contributions from emitter, stray light or measurement dark counts. We test our prediction under noise impacts by considering multiphoton effects of the displacement operator $D(\nu)=\exp(\nu a^{\dagger}-\nu^* a)$ with random amplitude $\nu$ on the emerging light. Thus, the density matrix deteriorated by the noise obtains
\begin{equation}
    \rho=\int \mathrm{d}^2\nu P(\nu) D(\nu)\rho_{-}D^{\dagger}(\nu)
    \label{noisyRho}
\end{equation}
with $P(\nu)$ being a probability density function identifying the stochastic processes affecting the amplitude $\nu$. We can analyse two typically relevant stochastic processes. In the first case only the phase of $\nu$ undergoes the fluctuation, i.e. $P_P(\nu)=\delta(|\nu|^2-\bar{n})/(2\pi)$. In the second case, we allow for $P_{BE}(\nu)=\exp\left\{-\left[(\nu+\nu^*)^2+(\nu-\nu^*)^2\right]/(4 \bar{n})\right\}/(\pi \bar{n})$, and therefore the amplitude $\nu$ randomly changes both the phase and its size. The model of the noise in (\ref{noisyRho}) with the probability density function $P_P(\nu)$ corresponds to the Poissonian noise and the model induced by the probability density function $P_{BE}(\nu)$ describes the Bose-Einstein noise. Appendix C presents how the Poissonian noise and the Bose-Einstein noise influence the manifestation of quantum non-Gaussianity for specific values of $\bar{n}$. In the region of very weak pumping power $|\alpha|^2 \ll 1$ and small mean number of noisy photons $\bar{n} \ll 1$, the probabilities obey the approximation
\begin{equation}
\begin{aligned}
         P_s &\approx \frac{\eta}{2}\\
        P_e &\approx \frac{\eta\left[(\eta-1)g^2+2\eta \gamma \kappa\right]^2}{g^4}|\alpha|^2+\eta \bar{n}
\end{aligned}
\end{equation}
for both the Poissonian noise and the thermal noise. Since the criterion becomes approximated by $P_s^3> P_e/4$ for the surpassed probability $P_e$, the quantum non-Gaussianity imposes an approximate condition
\begin{equation}
    |\alpha|^2+\bar{n} <\frac{\eta^2}{2},
    \label{approxM}
\end{equation}
which expresses the limitation on $|\alpha|^2$ involving the impacts of the noise. If $\bar{n} \geq \eta^2/2$ quantum non-Gaussianity is recognized for no values of $|\alpha|^2$. Let us recall that (\ref{noisyRho}) introduces a model of narrow-band noise, which contributes to the mode occupied by the light from the cavity.  A model of broad-band (incoherent) Poissonian noise, which contributes independently to many modes, gives rise to the less demanding approximate condition $2|\alpha|^2+\bar{n} <\eta^2$ with $\bar{n}$ being the mean number of noisy photons.

\section{Conclusion and Outlook}
To conclude, we proved that light emitted from a dispersively coupled atom or solid state emitter inside a cavity manifests quantum non-Gaussianity much more broadly under realistic conditions of current experiments. The only limitation for the recognition is a background noise deteriorating the measured light. High intensity of the pumping coherent beam, which prevents the recognition as well, can always be reduced with the cost of decreasing the overall rates. Simultaneously, this detection can be employed to expose the quantum non-Gaussianity of light from a quantum dot inside a micropillar cavity \cite{Najer2019} or a photonic crystal \cite{Lodahl2015}. Using more atoms in the cavity, future higher Fock state preparation can be evaluated using the existing hierarchy of quantum non-Gaussianity \cite{Lachman2019}.

\begin{acknowledgments}
R. F. and L. L. acknowledge project 21-13265X of the Czech Science Foundation. J. K. V. acknowledges the MSMT of the Czech Republic and Horizon 2020 Framework Programme (731473, project 8C20002 ShoQC). We acknolwedge the support of H2020 Spreading Excellence and Widening Participation (951737, NONGAUSS).
\end{acknowledgments}

\newpage
\appendix

\section{Appendix: Criterion of quantum non-Gaussianity}
The direct experimental recognition of quantum non-Gaussianity exploits the Hanbury Brown and Twiss configuration as depicted in Fig.~1 of the main text. It allows us to determine the probability $P_s$ of a click event on the detector SPAD$_1$ and the probability $P_e$ of double-click happening simultaneously on both the detectors SPAD$_1$ and SPAD$_2$. The criterion rejects directly the detectors' response on all the mixtures of Gaussian states in this specific scheme. It is represented by establishing a witness
\begin{equation}
    F_a(\rho)=P_s+a P_e,
\end{equation}
which exceeds a threshold value given for some freely chosen parameter $a$. The threshold value $F(a)$ corresponds to the maximum
\begin{equation}
    F(a)=\max_{\alpha,\xi} F_a(|\psi_G\rangle)
    \label{FaG}
\end{equation}
over the parameters $\alpha$ and $\xi$ identifying any pure Gaussian state $|\psi_G\rangle = S(\xi)D(\alpha)|0\rangle$ with $S(\xi)$ and $D(\alpha)$ being the squeezing and displacement operator. Importantly, the function $F(a)$ covers all mixtures of Gaussian states as well, i.e.
\begin{equation}
    F_a(\rho_G) \leq F(a)
    \label{mGCond}
\end{equation}
with $\rho_G$ being an incoherent mixture of pure states $|\psi_G \rangle$, and therefore violation of the inequality in (\ref{mGCond}) certifies the quantum non-Gaussianity. 

To express formulas for probabilities $P_s$ and $P_e$ that the state $|\psi_G\rangle$ exhibits, we introduce auxiliary probabilities $P_0$ and $P_{00}$ referring to the probability of no-click on SPAD$_1$ and to the probability of no-click on neither detector SPAD$_1$ and SPAD$_2$, respectively. These introduced probabilities obey the identities $P_s =1-P_0$ and $P_e =1-2P_0+P_{00}$. Then, the state $|\psi_G\rangle$ yields
\begin{equation}
\begin{aligned}
    P_0 &= 4\sqrt{\frac{V}{3V^2+10V+3}}e^{-\frac{|\alpha|^2\left[1+6V+V^2+(1-V^2)\cos 2\phi \right]}{2(3+V)(1+3V)}}\nonumber \\
    P_{00} &= 2\frac{\sqrt{V}}{1+V}e^{-\frac{|\alpha|^2\left[\cos^2 \phi +V \sin^2 \phi \right]}{1+V}},
\end{aligned}
\end{equation}
where $V=\exp(-2|\xi|)$ and $\phi$ is difference in phases of the squeezing $\xi$ and displacement $\alpha$. All these steps allow us to express $F_a(|\psi_G\rangle)$ as a function of the parameters $|\alpha|^2$, $V$ and $\phi$. Its maximum occurs when the parameters fulfill
\begin{equation}
\begin{aligned}
    \partial_{|\alpha|^2} P_s \partial_{\phi} P_e &=\partial_{|\alpha|^2} P_e \partial_{\phi} P_s\nonumber \\
    \partial_{|\alpha|^2} P_s \partial_{V} P_e &= \partial_{|\alpha|^2} P_e \partial_{V} P_s.
\end{aligned}
\end{equation}
These equations provide a solution $\phi=0$ and $|\alpha|^2=(V+3)(1-V^2)/\left[2V(1+3V)\right]$ with $V \in \left(0,1 \right]$. Thus, all mixtures of Gaussian states $\rho_G$ satisfy
\begin{equation}
\begin{aligned}
    F_a(\rho_G) &\ \leq \max_V \left\{ 1-P_0(V)\right. \nonumber\\ 
     &+ \left. a\left[1-2P_0(V)+P_{00}(V)\right]\right\}
\end{aligned}
     \label{FaExp}
\end{equation}
with 
\begin{equation}
\begin{aligned}
     P_0(V)&= 4 e^\frac{-1+V^2}{2V(1+3V)}\sqrt{\frac{V}{3+10V+3V^2}} \\
    P_{00}(V)&= 2 e^\frac{(-1+V)(1+3V)}{2V(1+3V)}\frac{\sqrt{V}}{1+V}.
\end{aligned}
\end{equation}
Recall, the quantum non-Gaussianity is witnessed by such a parameter $a$ that implies surpassing the right hand side of (\ref{FaExp}) by $F_a$ of a measured state. This $a$ can be always found if the measured state yields probabilities above a threshold  in the space of probabilities $P_s$ and $P_e$. Its parametric form gains
\begin{equation}
\begin{aligned}
    P_s&=1-P_0(V) \\
    P_e&=1-2P_0(V)+P_{00}(V),
\end{aligned}
    \label{SM:thresNonG}
\end{equation}
where $V \in \left(0,1\right]$ plays the role of a parameter.

\section{Appendix: A single atom dispersively coupled  to a cavity}
Let pulsed laser light enter a cavity where an atom with energy levels forming the $\Lambda$-system is coupled to the cavity mode. Under assumptions mentioned in \cite{Wang2005}, the heralded light leaving the cavity approaches the density matrix \cite{Hacker2019}
\begin{equation}
\begin{aligned}
     &\ \rho_{-}=\frac{1}{2(1-e^{-2\eta \vert \alpha \vert^2})}\left[\vert \alpha_e \rangle \langle \alpha_e \vert+\vert \alpha_g \rangle \langle \alpha_g \vert \right. \\ 
    &\ \left. -e^{-2(1-\eta)\eta \vert \alpha \vert^2}\left(\vert \alpha_e \rangle \langle \alpha_g \vert+\vert \alpha_g \rangle \langle \alpha_e \vert\right)\right],
\end{aligned}
    \label{catDMSM}
\end{equation}
where $|\alpha_e \rangle$ and $|\alpha_g \rangle$ are coherent states with the amplitudes
\begin{equation}
    \alpha_g=\frac{\kappa_l-\kappa_r}{\kappa_l+\kappa_r} \alpha, \  \alpha_e=\frac{g^2+(\kappa_l-\kappa_r)\gamma}{g^2+(\kappa_l+\kappa_r)\gamma}\alpha.
    \label{alfGalfESM}
\end{equation}
with $\alpha$ being the amplitude of the incoming coherent state. The parameter $\eta$ in (\ref{catDMSM}) is defined by the identity
\begin{equation}
    \eta |\alpha |=\frac{1}{2}|\alpha_e-\alpha_g|,
    \label{diff}
\end{equation}
which gives rise to
\begin{equation}
    \eta=\frac{\kappa_r}{\kappa_l+\kappa_r}\frac{g^2}{g^2+\gamma (\kappa_l+\kappa_r)}.
    \label{etaSM}
\end{equation}
It shows how the cavity parameters reduce the difference (\ref{diff}). A cavity operating in the strong coupling regime when $g \gg \gamma$ and $g \gg \kappa_r \gg \kappa_l$ allows us to approach $\eta=1$. Beyond that regime, the cumulative parameter $\eta$ is always reduced below one, which affects the purity of the state $\rho_-$.

\subsection{Wigner function}
The Wigner function $W(\alpha,\alpha^*)$ enables a recognition of the quantum non-Gaussianity when it exhibits negativity. To determine its value for the states $\rho_{-}$, we consider the relation
\begin{equation}
    W(\beta,\beta^*)=\frac{1}{2\pi}\langle D^{\dagger}(\beta)(-1)^{a^\dagger a}D(\beta)\rangle,
\end{equation}
where $a^{\dagger}$ and $a$ are creation and annihilation operator and $D(\beta)=\exp \left(\beta a^{\dagger}-\beta^* a\right)$ is the displacement operator \cite{Royer1977}. Let us introduce the state
\begin{equation}
    \widetilde{\rho}_-=D\left[-(\alpha_e+\alpha_g)/2\right]\rho_- D^{\dagger}\left[-(\alpha_e+\alpha_g)/2\right].
    \label{tilderho}
\end{equation}
The displacement operator in (\ref{tilderho}) shifts the state $\rho_-$ such that the negative region (if there is any) is at the origin. Thus, employing the identities (\ref{catDM}) and (\ref{alfGalfESM}) yields
\begin{equation}
\begin{aligned}
     &\ \widetilde{\rho}_{-}=\\
    &\ \frac{1}{2(1-e^{-2\eta \vert \alpha \vert^2})}\left[\vert \eta\alpha \rangle \langle \eta\alpha \vert+\vert -\eta\alpha \rangle \langle \eta\alpha \vert \right. \\
    &- \left. e^{-2(1-\eta)\eta \vert \alpha \vert^2}\left(\vert \eta\alpha \rangle \langle -\eta\alpha \vert+\vert -\eta\alpha \rangle \langle -\eta\alpha \vert\right)\right].
\end{aligned}
    \label{catDMSifhted}
\end{equation}
The state $\widetilde{\rho}_{-}$ is formally identical to the state $|\psi\rangle \propto (|\alpha\rangle-|-\alpha\rangle)$ that undergoes losses quantified by the transmission $\eta$. Due to that, the negativity remains only for $\eta>1/2$ \cite{Hudson1974}.

\subsection{Criterion from Ref.\cite{Genoni2013}}
\begin{figure}
    \centering
    \includegraphics[width=1\linewidth]{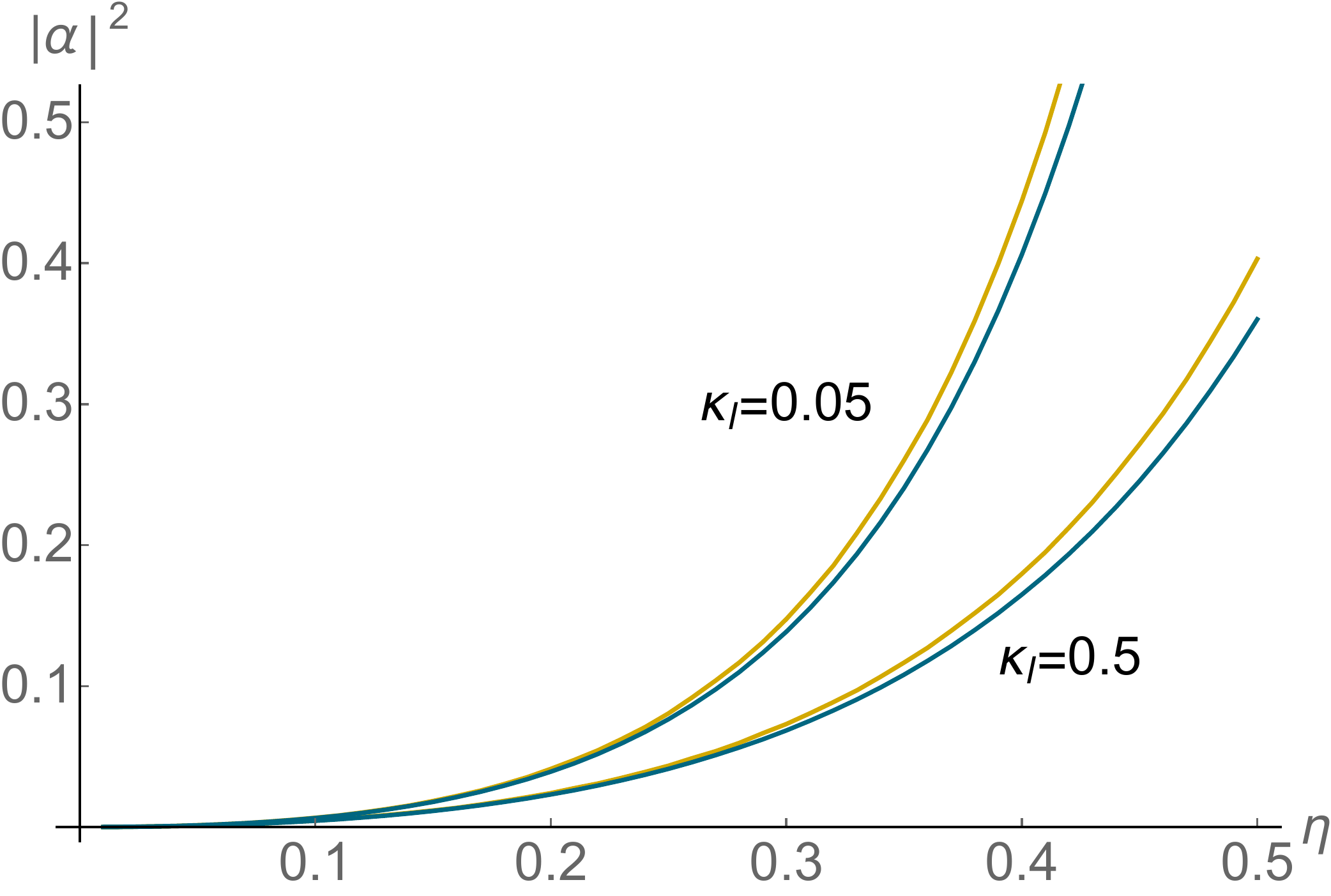}
    \caption{Comparing thresholds in terms of intensity of pumping $|\alpha|^2$ and the parameter $\eta$ that enable certifying the quantum non-Gaussianity of the state  (\ref{catDMSM}). The thresholds imply from obeying the criterion (\ref{PsPeThPNRD}) in the main text (yellow) or from satisfying the criterion (\ref{critWig}) (blue). The spontaneous emission rate $\gamma$ obtains $\gamma=0.32$ and the values of the parameter $\kappa_l$ are fixed to $\kappa_l=0.05$ and $\kappa_l=0.5$.}
    \label{fig:compareCrit}
\end{figure}

The certification of the quantum non-Gaussianity for states having completely positive Wigner function is also possible employing the criterion combining the value of the Wigner function at origin $W(0,0)=\mbox{Tr}\left[\rho (-1)^{a^\dagger a}\right]$ with the mean number of photons $\bar{N}=\langle a^{\dagger} a\rangle$. Explicitly, the criterion recognizes the quantum non-Gaussianity when
\begin{equation}
    W(0,0)<\frac{1}{2\pi}e^{-2\bar{N}(1+\bar{N})}.
    \label{critWig}
\end{equation}
The condition converges to the requirement on the negativity for large $\bar{N}$. However, the quantum non-Gaussianity can manifest itself even for $W(0,0)$ arbitrarily close to $\frac{1}{2\pi}$ (a value achieved by the vacuum) when $\bar{N}$ is sufficiently suppressed. The criterion (\ref{critWig}) applied on the state (\ref{catDMSM}) reveals that the quantum non-Gaussianity can manifests itself for arbitrarily small $\eta>0$. Fig.~\ref{fig:compareCrit} compares the criterion (\ref{PsPeThPNRD}) in the main text with the criterion (\ref{critWig}) with respect to their capability to recognize the quantum non-Gaussianity of the state (\ref{catDMSM}). Both criteria impose  similar demands on the state (\ref{catDMSM}). Let us emphasise, however, that the criteria are very different concerning the requirement for detection. Criterion (\ref{PsPeThPNRD}) requires only faithful resolution of zero, single and more photons, whereas (\ref{critWig}) needs reliable and faithful photon-number resolving detection. 

\section{Appendix: Effects of the noise}
The density matrix (\ref{catDM}) represents an emitted light, which is not affected by noise. To include the noise contributions in the description, we allow for a scenario when the light is influenced by the action of a displacement operator with a fluctuating amplitude. Thus, the noise affects the density matrix according to
\begin{equation}
    \rho=\int \mathrm{d}^2 \beta P(\beta)D(\beta)\rho_{-}D^{\dagger}(\beta),
    \label{rhoNoise}
\end{equation}
where $D(\beta)=\exp(\beta a^{\dagger}-\beta^* a)$ is the displacement operator. The amplitude $\beta$ undergoes random fluctuations that the probability density function $P(\beta)$ determines. Further, we assume two fundamental stochastic processes for which the probability density functions yield
\begin{equation}
\begin{aligned}
      P_P(\beta)&=\frac{1}{2\pi}\delta \left(\vert \beta \vert- \sqrt{\bar{n}}\right) \\
      P_{BE}(\beta)&=\frac{1}{\pi \bar{n}}e^{-\frac{\vert \beta \vert^2}{\bar{n}}},
\end{aligned}
      \label{PDFs}
\end{equation}
where $\bar{n}$ is mean number of the noisy photons.
When $P_P(\beta)$ is taken into account, the amplitude $\beta$ experiences random fluctuation in phase only. On the contrary, $P_{BE}(\beta)$ represents fluctuation in both the phase and the intensity $\vert \beta \vert^2$. Let us call the noise identified by the probability density functions $P_P(\beta)$ Poissonian noise and the noise given by $P_{BE}(\beta)$ Bose-Einstein noise according to the distribution of noisy photons that are produced when the stochastic displacement affects the vacuum states.

\begin{figure*}[ht!]
\centering
\includegraphics[width=1\linewidth]{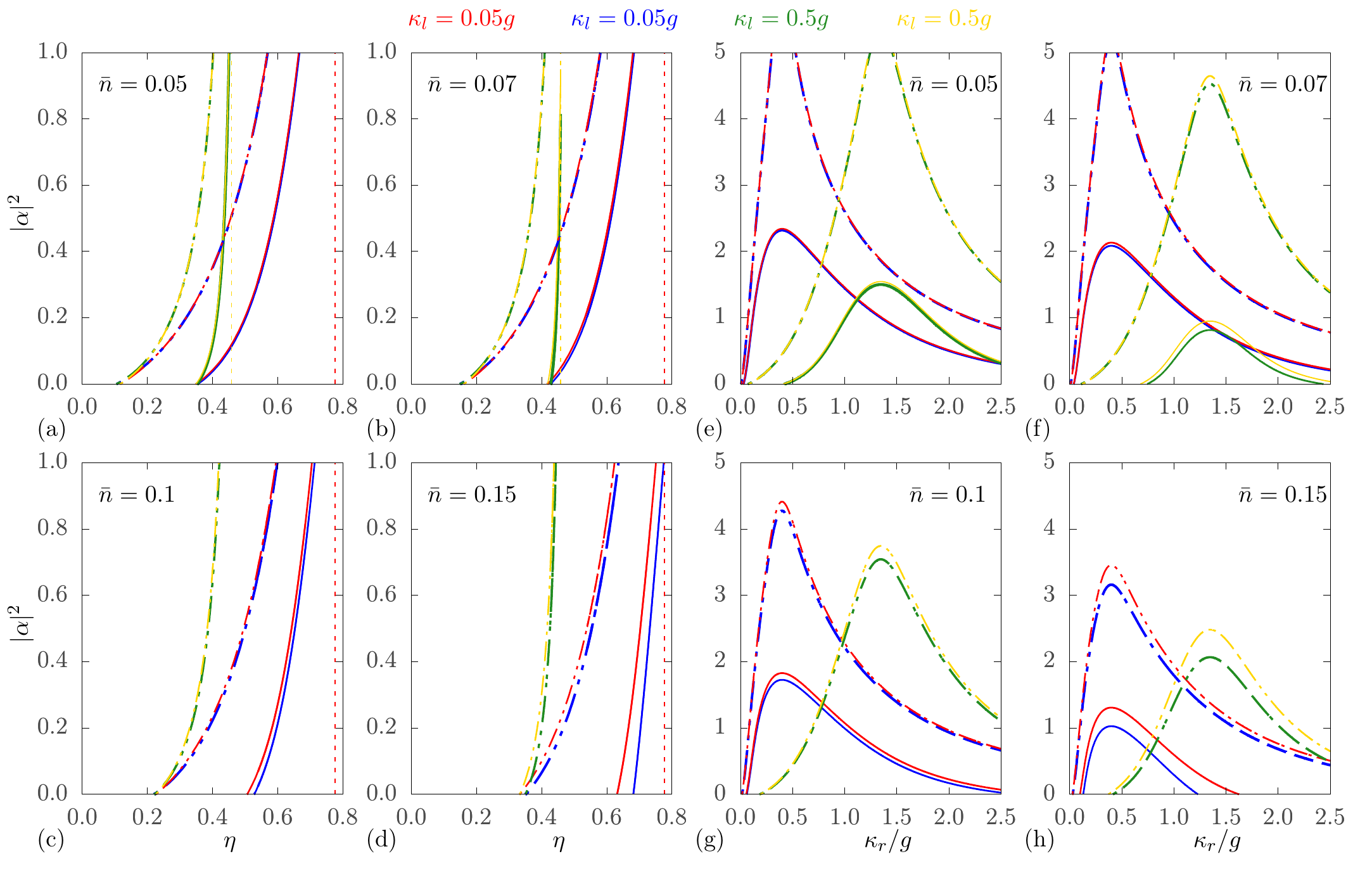}
    \caption{Thresholds of the quantum non-Gaussianity and nonclassicality of noisy states defined by (\ref{rhoNoise}) and  (\ref{PDFs}). The thresholds are shown in terms of pumping intensity $|\alpha|^2$ and either cumulative parameter $\eta$ (\emph{a})-(\emph{d}) or the parameter $\kappa_r$ normalized to $g$ (\emph{e})-(\emph{h}). The quantum non-Gaussianity and nonclassicality are certified for $|\alpha|^2$ below a respective threshold. Each sub-figure allows for different mean number of noisy photons $\bar{n}$. Whereas the blue and green solid lines represent the Poissonian noise deteriorating the light, the red and golden solid lines correspond to the Bose-Einstein noise. Also, the colors differentiate the parameter $\kappa_l$, which obtain either $\kappa_l=0.05 g$ (red and blue) or $\kappa_l=0.5 g$ (green and golden) as labeled above the plots. The spontaneous emission rate $\gamma$ is set to $\gamma=0.32 g$ for each sub-figure. Finally, a line of a given color is plotted as solid and dashed, which allows comparison of the thresholds for the quantum non-Gaussianity (solid) with thresholds for the nonclassicality (dashed). Sub-figures  (\emph{c}), (\emph{d}) and (\emph{g}), (\emph{h}) do not present the solid golden and green lines since they depict thresholds of states where the noise contributions effects lost of the quantum non-Gaussianity when $\kappa_l=0.5$.  In sub-figures (\emph{a})-(\emph{d}), the horizontal red and golden dashed lines represent the physical boundary for $\eta$  when $\kappa_l=0.05 g$ and $\kappa_l=0.5 g$, respectively.}
\label{fig:fig5}
\end{figure*}

To obtain formulas for the click probabilities, let us deal with the state $\rho_{-,\beta}=D(\beta)\rho_{-}D^{\dagger}(\beta)$ before considering the stochastic processes. Its density matrix works out to
\begin{equation}
\begin{aligned}
   &\ \rho_{-,\beta}=\frac{1}{2(1-e^{-2\eta |\alpha|^2})}\left\{|\bar{\alpha}_{e,\beta}\rangle\langle \bar{\alpha}_{e,\beta}|+|\bar{\alpha}_{g,\beta}\rangle\langle \bar{\alpha}_{g,\beta}|\right.\\
    &\ -e^{-2(1-\eta)\eta|\alpha|^2} \times \left[e^{\eta (\beta \alpha^*-\beta^* \alpha)}|\bar{\alpha}_{e,\beta}\rangle\langle \bar{\alpha}_{g,\beta}| \right. \\
    &\ \left.\left.+e^{\eta (\beta^* \alpha-\beta \alpha^*)}|\bar{\alpha}_{g,\beta}\rangle\langle \bar{\alpha}_{e,\beta}|\right]\right\},
\end{aligned}
\label{incRho}
\end{equation}
where $\bar{\alpha}_{e,\beta}=\alpha_e+\beta$ and $\bar{\alpha}_{g,\beta}=\alpha_g+\beta$.
Let the state $\rho_{-,\beta}$ go through a balanced BS. Outgoing state $\bar{\rho}_{-,\beta}$ in the transmitted mode is given by
\begin{equation}
\begin{aligned}
    &\ \bar{\rho}_{-,\beta}=\\
    &\ \frac{1}{2(1-e^{-2\eta |\alpha|^2})}\left\{|\bar{\alpha}_{e,\beta}/\sqrt{2}\rangle\langle \bar{\alpha}_{e,\beta}/\sqrt{2}|\right. \\
    &\ +|\bar{\alpha}_{g,\beta}/\sqrt{2}\rangle\langle \bar{\alpha}_{g,\beta}/\sqrt{2}|\\ 
    &\ -e^{-2(1-\eta)\eta|\alpha|^2-\eta^2 |\alpha|^2+\left[\beta^*(\alpha_{g}+\alpha_{e})+\beta(\alpha_{g}^*+\alpha_{e}^*)\right]/4} \\
    &\ \left. \times \left[e^{(\alpha^*_e \beta+\alpha_g \beta^*)/2+\eta (\beta \alpha^*-\beta^* \alpha)}|\bar{\alpha}_{e,\beta}/\sqrt{2}\rangle\langle \bar{\alpha}_{g,\beta}/\sqrt{2}|\right. \right. \\
    &\ +e^{(\alpha_e \beta^*+\alpha_g^* \beta)/2+\eta (\beta^* \alpha-\beta \alpha^*)}\\
    &\ \left. \left. \times |\bar{\alpha}_{g,\beta}/\sqrt{2}\rangle\langle \bar{\alpha}_{e,\beta}/\sqrt{2}|\right]\right \}.
\end{aligned}
\label{ougRho}
\end{equation}
The density matrices of the incident state (\ref{incRho}) and transmitted state (\ref{ougRho}) allows us to determine how $\beta$ influences the click probability $P_s(\beta)=1-\langle 0|\bar{\rho}_{-,\beta}|0\rangle$ and double click $P_e(\beta)=1-2\langle 0|\bar{\rho}_{-,\beta}|0\rangle+\langle 0|\rho_{-,\beta}|0\rangle$. Thus, the click probability $P_s$ and double click probability $P_c$ of the noisy state (\ref{rhoNoise}) read
\begin{equation}
\begin{aligned}
     P_s=\int \mathrm{d}^2 \beta P(\beta) P_s(\beta) \\
    P_e=\int \mathrm{d}^2 \beta P(\beta) P_e(\beta)
\end{aligned}
\end{equation}
with $P(\beta)$ being some probability density function from (\ref{PDFs}). We provide only the approximate formulas for $P_s$ and $P_e$ appropriate for states with $|\alpha|^2 \ll 1$ and $\bar{n}\ll 1$. Under these assumptions, they get simplified to
\begin{equation}
\begin{aligned}
     P_s&=\frac{\eta}{2}\\
    P_e&=\frac{\eta\left[(\eta-1)g^2+2\eta \gamma \kappa\right]^2}{g^4}|\alpha|^2+\eta \bar{n},
\end{aligned}
\label{approx1}
\end{equation}
for both the considered noise. Fig.~\ref{fig:fig5} depicts when the noise contribution enables recognition of the quantum non-Gaussianity for some specific settings of the cavity parameters.

The model of noisy states (\ref{rhoNoise}) can be modified to a model in which the noise contributes as independent background noise, i.e. the density matrix reads $\rho=\rho_-\otimes \rho_{\bar{n}}$, where $\rho_{\bar{n}}$ obeys the Poissonian distribution of photons with mean number of photons $\bar{n}$. In this case, the click probabilities in the limit of $|\alpha|^2 \ll 1$ and $\bar{n}\ll 1$ works out to be
\begin{equation}
\begin{aligned}
    P_s&=\frac{\eta}{2}\\
    P_c&=\frac{\eta\left[(\eta-1)g^2+2\eta \gamma \kappa\right]^2}{g^4}|\alpha|^2+\frac{\eta \bar{n}}{2}.
\end{aligned}
\label{approx2}
\end{equation}
Comparing (\ref{approx1}) with (\ref{approx2}) suggests that the background noise has less impacts on the probability $P_c$. Consequently, the quantum non-Gaussianity is more sensitive to contributions of noise in (\ref{rhoNoise}) than to the contributions of the background noise.
 
 \bibliographystyle{abbrvnat}
\bibliography{references.bib}
\end{document}